\documentclass[11pt]{article}
\usepackage{amsfonts}
\usepackage{latexsym}
\usepackage{amsmath,amssymb}
\usepackage{verbatim}
\usepackage{setspace}
\usepackage{color}
\usepackage{physics}
\usepackage{tikz}
\usepackage{mathdots}
\usepackage{yhmath}
\usepackage{cancel}
\usepackage{color}
\usepackage{siunitx}
\usepackage{mathrsfs}  
\usepackage{array}
\usepackage{multirow}
\usepackage{amssymb}
\usepackage{gensymb}
\usepackage{tabularx}
\usepackage{hyperref}
\usepackage{booktabs}
\numberwithin{figure}{section}
\usetikzlibrary{fadings}
\usetikzlibrary{patterns}
\usetikzlibrary{shadows.blur}
\usetikzlibrary{shapes}

\usepackage[textheight=9in, textwidth=6.5in, letterpaper]{geometry}

\numberwithin{equation}{section}

\renewcommand{\ip}{${\mathcal I}^+$}

 \def\p{\partial}
 
 \def\bz{{\bar z}}
 
\def\0{{(0)}}
\def\1{{(1)}}
\def\2{{(2)}}

\def\n{\nabla}

\def\<{\langle }
\def\>{\rangle }
\def\[{\left[}
\def\]{\right]}
\def\bw{{\bar w}}

\newcommand{\bea}{\begin{eqnarray}}
\newcommand{\eea}{\end{eqnarray}}
\newcommand{\be}{\begin{equation}}
\newcommand{\ee}{\end{equation}}
\newcommand{\ba}{\begin{align}}
\newcommand{\ea}{\end{align}}

\renewcommand{\O}{\mathcal{O}}

\renewcommand{\epsilon}{\varepsilon}

   \makeatletter
  \let\over=\@@over \let\overwithdelims=\@@overwithdelims
  \let\atop=\@@atop \let\atopwithdelims=\@@atopwithdelims
  \let\above=\@@above \let\abovewithdelims=\@@abovewithdelims
\renewcommand\section{\@startsection {section}{1}{\z@}%
                                   {-3.5ex \@plus -1ex \@minus -.2ex}
                                   {2.3ex \@plus.2ex}%
                                   {\normalfont\large\bfseries}}

\renewcommand\subsection{\@startsection{subsection}{2}{\z@}%
                                     {-3.25ex\@plus -1ex \@minus -.2ex}%
                                     {1.5ex \@plus .2ex}%
                                     {\normalfont\bfseries}}

\begin{document}
\unitlength = 1mm
\ \\
\vskip1cm
\begin{center}

{ \LARGE {\textsc{Celestial Gluon Amplitudes from the Outside In}}}

\vspace{0.8cm}
Walker Melton{$^{*}$}, Sruthi A. Narayanan{$^{\dagger}$}
\vspace{1cm}

{$^*$\it  Center for the Fundamental Laws of Nature, Harvard University,\\
Cambridge, MA 02138, USA} \\
{$^\dagger$\it Perimeter Institute for Theoretical Physics, \\
Waterloo, ON N2L 2Y5, Canada}

\vspace{0.8cm}

\begin{abstract}
We show that, given a two-dimensional realization of the celestial OPE in self-dual Yang-Mills, we can find a scalar source around which scattering amplitudes replicate correlation functions computed from the 2D `gluon' operators in a limit where a dynamic massless scalar decouples. We derive conditions on the two-dimensional three-point correlation function so that such a source exists and give two particular examples of this construction, one in which gluons are constructed from vertex operators in the semiclassical limit of Liouville theory and another in which the soft gluons arise from generalized free fields. Finally, we identify a bulk dual to the level of the boundary Kac-Moody algebra and discuss moving beyond the decoupling limit. 
 \end{abstract}
\vspace{0.5cm}

\vspace{1.0cm}

\end{center}

\pagestyle{empty}
\pagestyle{plain}
\newpage
\tableofcontents
\def\gzz{\gamma_{z\bz}}
\def\vx{{\vec x}}
\def\p{\partial}
\def\po{$\cal P_O$}
\def\cN{{\cal N}_\rho^2 }
\def\N{${\cal N}_\rho^2 ~~$}
\def\G{\Gamma}
\def\a{{\alpha}}
\def\b{{\beta}}
\def\g{\gamma}
\def\ch{{\cal H}^+}
\def\hf{{\cal H}}
\def\Im{\mathrm{Im\ }}
\def\bpd{\bar{\partial}}
\def\hbh{{\cal H}_{\rm BH}}
\def\hout{{\cal H}_{\rm OUT}}
\def\ss{\Sigma_S}
\def\D{{\rm \Delta}}
\def \bw {{\bar w}}
\def \bz {{\bar z}}
\def\cS{{\cal S}}
\def\l{\lambda }
\def\d{{\delta}}
\def\n{{\rm SC}}
\def\ip{{\rm cft}}
\def\RR{\mathbb{K}}
\def\i{i^\prime}
\def\scri{\mathcal{I}}
\def\A{\mathcal{A}}
\def\zb{\bar{z}}
\def\adz{AdS$_3/\mathbb{Z}$}
\def\sll{$SL(2, \mathbb{R})_L$}
\def\slr{$SL(2, \mathbb{R})_R$}
\renewcommand{\aa}[1]{\left\langle#1\right\rangle}
\pagenumbering{arabic}
\section{Introduction}
Celestial holography posits a holographic duality between a theory of quantum gravity and a co-dimension two `celestial' conformal field theory that lives on the  celestial sphere. Correlation functions in this putative celestial CFT calculate the $\mathcal{S}$-matrix of the bulk theory, which, as it is defined from data living on null infinity, is a naturally holographic observable. While few examples of these dual theories are known, one can calculate correlation functions in the dual celestial CFT from bulk scattering amplitudes by putting the external particles in a basis of wavefunctions that transform as conformal primaries under Lorentz transformations, which act as global conformal transformations on the celestial sphere~\cite{Pasterski:2016qvg, Pasterski:2017kqt}. 

To date, majority of the progress in celestial holography has been a result of working from the inside out: that is, one identifies a property of bulk amplitudes that, when transformed to a conformal basis, becomes some interesting property of the putative celestial conformal field theory. Thus far, this has been a useful technique for studying various aspects of celestial holography. For instance, by Mellin transforming collinear splitting functions in Yang-Mills and gravity, one can uncover the singular terms in the OPEs between gluons and gravitons~\cite{Guevara:2021abz}. Additionally, Mellin transforming universal soft theorems indicates that celestial CFTs that are dual to gravity contain currents whose OPEs with other particles encode universal soft theorems~\cite{Pate:2019lpp, Puhm:2019zbl}. Combining these results lead to the discovery of the universal soft $w_{1+\infty}$ and $S$-algebras in gravity and gauge theory, respectively~\cite{Himwich:2021dau, Strominger:2021mtt}. 

While this has been a useful tool for studying celestial CFTs dual to simple bulk theories, one would like to have examples where the boundary theory has an independent definition.  While one might hope to find a bulk theory whose scattering amplitudes allow one to read off the 2D CFT that computes them, celestial amplitudes, especially those of massless particles where the holographic dictionary is most well understood, tend to have unusual properties. For example, due to bulk translation invariance, low-point celestial amplitudes of massless particles are distributional~\cite{Law:2019glh, Pasterski:2017ylz}. Understanding how these distributional correlation functions could arise from a two-dimensional conformal field theory, where correlation functions are generally smooth, is an important issue~\cite{Melton:2023bjw}. Nevertheless, we will not address this here. Instead, we will break translation invariance by adding an explicit source term for a scalar field. This allows celestial amplitudes of massless particles to be non-singular on the celestial sphere~\cite{Casali:2022fro, Melton:2022def, Fan:2022ele, Gonzo:2022tjm, Taylor:2023bzj}. 

Scattering amplitudes with non-trivial backgrounds have appeared recently in proposals for several explicit celestial dualities. In~\cite{Costello:2022jpg,Costello:2023hmi}, Costello, Paquette and Sharma argued that a particular four-dimensional WZW model coupled to an unusual scalar gravity theory in a four-dimensional asymptotically flat  spacetime was dual to the large $N$ limit of a simple 2D chiral algebra. In~\cite{Stieberger:2022zyk}, it was argued that amplitudes calculated from a simple boundary theory consisting of a centerless Kac-Moody algebra and the semiclassical limit of Liouville theory matched tree-level MHV amplitudes computed around a $\delta$-function source for a canonical scalar field.  In~\cite{Stieberger:2023yml}, they suggest that this duality might persist beyond tree level.

Interestingly, introducing source terms centrally extends the soft algebra and allows soft currents to create states of finite norm under the familiar CFT inner product. Additionally, while the source does not deform the singular part of the gluon-gluon-gluon OPE, the relationship between the three- and two-point functions implies that finite terms in the gluon OPE depend on ratios of different terms in a Taylor expansion of the background~\cite{Melton:2022def}. This suggests that surprisingly different boundary theories might be dual to surprisingly similar bulk theories with different backgrounds.

In this work, we attempt to use the general features of celestial amplitudes from the ``inside out" perspective to find some toy examples of 2D theories that may have a bulk dual. We focus on realizations of the soft algebra for self-dual Yang-Mills (SDYM) theory where the gluon operator factors into a Kac-Moody current that is the same for all gluons and encodes the color structure of the theory and a scalar primary operator that provides the correct scaling dimension. We then show that, for some different realizations of this scalar theory, we can identify a scalar background such that connected tree-level all $+$ amplitudes in SDYM coupled with a $\phi F^2$ interaction to a neutral scalar match those computed from the toy two-dimensional theory in a limit where the coupling to the scalar vanishes and the strength of the background diverges. 

This paper proceeds as follows. After reviewing celestial holography and self-dual Yang-Mills theory in Section~\ref{sec:prelim}, Section~\ref{sec:algdual} defines an ansatz for two-dimensional realizations of the gluon-gluon algebra in which the gluon splits into a universal Kac-Moody current and a scalar operator, possibly with a summed-over internal index. We explain how to obtain correlators from this in a particular decoupling limit and derive a condition on the three-point function of the scalar operator such that a scalar source exists and amplitudes around that source match the three-point function computed from the candidate boundary theory. In Sections~\ref{sec:vertex} and~\ref{sec:gff}, we give two examples of these boundary theories and a bulk source whose scattering amplitudes match the boundary correlation functions. Finally, in Section~\ref{sec:scoup} we discuss moving beyond the decoupling limit and identify a bulk coupling constant dual to the level of the universal Kac-Moody currents in the 2D dual.
\section{Preliminaries}
\label{sec:prelim}
In this section, we will review the origins of the celestial OPE as a chiral collinear limit as well as the tree-level $\mathcal{S}$-matrix of SDYM coupled to a scalar source. 
\subsection{Celestial Amplitudes}
Celestial amplitudes provide a way to interpret four-dimensional scattering amplitudes as correlation functions in a two-dimensional celestial CFT. To see this, one must first appropriately parametrize the spacetime. In Cartesian coordinates, flat Minkowski space has metric
\begin{equation}
    ds^2 = -dx_0^2+dx_1^2+dx_2^2+dx_3^2
\end{equation}
in $(3,1)$ signature.\footnote{Most of the literature treats $z,\bar{z}$ independently, either as real and independent variables or by separately complexifying them. While we work in $(3,1)$ we can analytically continue to $(2,2)$ to make connections to those results.} A null vector $q^\mu$ can be parametrized by 
\begin{equation}
\begin{split}
    q^\mu(\omega,z,\zb) &= \eta\omega \hat{q}^\mu(z,\zb) \\
    \hat{q}^\mu(z,\zb) &= (1+z\zb,z+\zb,-i(z-\zb),1-z\zb)
\end{split}
\end{equation}
where $\omega$ is a frequency, $\eta = +1\ (-1)$ for outgoing (incoming) particles, and $(z,\zb)$ labels a coordinate on the celestial sphere $\mathcal{CS}$. Because we are working in (3,1) signature, $z$ and $\zb$ are complex conjugates of one another. Under Lorentz transformations, $z, \zb$ transform under M\"{o}bius transformations. Using the standard basis for Pauli matrices $\sigma^\mu$, we can write any momentum as $q_{\alpha\dot{\alpha}} = (q\cdot \sigma)_{\alpha\dot{\alpha}} = \lambda_{\alpha}\tilde{\lambda}_{\dot{\alpha}}$ where these spinors then define the following spinor helicity variables with brackets
\begin{equation}
\langle ij\rangle = -2\eta_i\eta_j\sqrt{\omega_i\omega_j}z_{ij}, \ \ [ij] = -2\sqrt{\omega_i\omega_j}\bar{z}_{ij}
\end{equation}
in the usual way, such that $\langle ij\rangle[ij] = -2q(\omega_i,z_i,\bar{z}_i)\cdot q(\omega_j,z_j,\bar{z}_j)$. 

Celestial amplitudes can be found by computing scattering amplitudes in a basis that diagonalizes boosts rather than translations. This is referred to as the celestial basis or the conformal primary basis. For massless external particles, these can be found by Mellin transforming the momentum space scattering amplitude with respect to the frequency $\omega$
\begin{equation}
    \begin{split}
        \A\left(1^{\eta_1}_{\Delta_1,a_1}\cdots n^{\eta_n}_{\Delta_n,a_n}\right) &= \int_0^\infty \prod_{j=1}^nd\omega_j\omega_j^{\Delta_j-1}A_{a_1\ldots a_n}(\eta_1\omega_1\hat{q}(z_1,\zb_1),\ldots,\eta_n\omega_n\hat{q}(z_n,\zb_n))
    \end{split}
\end{equation}
where $\Delta_n$ is the conformal weight of the external particles and $a_n$ is any other label, such as helicity, color, or species. Lorentz invariance of the momentum space scattering amplitude $A$ implies that the celestial amplitude $\A$ transforms as a correlation function of $n$ conformal primaries of weights $\Delta_i$ and spins given by the bulk helicities of the particles. As such, when it is convenient we will write these amplitudes as correlation functions 
\begin{equation}
    \begin{split}
        \A\left(1^{\eta_1}_{\Delta_1,a_1},\ldots n^{\eta_n}_{\Delta_n,a_n}\right) = \left\langle \O^{\eta_1}_{\Delta_1,a_1}\cdots \O^{\eta_n}_{\Delta_n,a_n}\right\rangle .
    \end{split}
\end{equation} 

\subsection{Celestial OPEs and Collinear Singularities}
In traditional conformal field theory one usually says the theory is specified by the spectrum, i.e the conformal dimensions of the operators, and their operator product expansion which expands a product of two local operators as a sum over a basis of single operators. In a conventional CFT, the existence of the OPE is often related to the state-operator correspondence, but it can also be studied by examining coincident limits of CFT correlation functions. 

For celestial amplitudes, taking $z_1 \to z_2$ for two celestial insertions is equivalent to taking $\langle 12 \rangle \to 0$ in the bulk scattering amplitude.\footnote{Note that this is a chiral collinear limit that differs from the standard collinear limit where $z_{12}, \zb_{12} \to 0$.} In this limit, gauge and gravitational amplitudes exhibit a universal factorization behavior where\footnote{Similar collinear factorizations exist for general helicity scattering amplitudes in full Yang-Mills theory.}
\begin{equation}
    \A_{n,++\ldots} = \mathrm{Split}^{-}_{\ ++}(p_1,p_2)\A_{(n-1),+\ldots}(p_1+p_2,p_3,\ldots).
\end{equation}
The singular term in the OPE arises from diagrams where the particles that are becoming collinear are attached to the same three-point vertex. As $z_{12} \to 0$, the internal propagator goes on-shell and leads to the collinear divergence. Mellin transforming this leading splitting function leads to the leading term in the celestial OPE. For gluons, this gives the OPE\footnote{In some cases where the soft algebra fails to satisfy the Jacobi identity, there can be another singular term in the OPE \cite{Ball:2023sdz}. We do not consider this case.}
\begin{equation}\label{eq:oggluonOPE}
    \O^{+a}_{\Delta_1}(z_1,\zb_1)\O^{+b}_{\Delta_2}(z_2,\zb_2) = -\frac{if^{ab}_{\ \ c}}{z_{12}}\sum_{m=0}^\infty \frac{\Gamma(\Delta_1+m-1)\Gamma(\Delta_2-1)}{m!\Gamma(\Delta_1+\Delta_2+m-2)}\zb_{12}^m\bar{\partial}_2^m\O^{+c}_{\Delta_1+\Delta_2-1}(z_2,\zb_2) + \O(z_{12}^0).
\end{equation}
where $f^{ab}_{\ \ c}$ are the usual structure constants in Yang-Mills. A similar story exists in gravity that we do not reproduce here~\cite{Guevara:2021abz}.

\subsection{Conformally Soft Theorems}
In momentum space, gauge and gravitational amplitudes exhibit universal behavior in the limit where one of the external particles become soft. In the dual celestial CFT, these soft theorems govern divergences in the celestial amplitudes of gluons as one of the external weights goes to specific integer values, $\Delta = 1, 0, \ldots$. As such, one can define the soft gluon operators
\begin{equation}
    R^{k,a}(z,\zb) = \lim_{\epsilon\to 0}\epsilon\O^{+a}_{k+\epsilon}(z,\zb),\ k = 1, 0, -1,\ldots 
\end{equation}
When inserted into correlation functions, these soft operators give finite correlation functions as the vanishing $\epsilon$ cancels a divergence arising from the small $\omega$ region of the Mellin transform. Taking the double soft limit of the gluon-gluon OPE~\eqref{eq:oggluonOPE} reveals that the OPE of soft gluons takes the form \cite{Guevara:2021abz}
\begin{equation}
    R^{k,a}(z_1,\zb_1)R^{\ell,b}(z_2,\zb_2) \sim -\frac{if^{ab}_{\ \ c}}{z_{12}}\sum_{j=0}^{1-k}\frac{(2-k-\ell-j)!}{(1-k-j)!(1-\ell)!j!}\zb_{12}^j\bpd^jR^{k+\ell-1,c}(z_2,\zb_2).
\end{equation}
Expanding in modes\footnote{Note that this is just the standard CFT mode expansion of operators where we have performed a convenient rescaling of the mode coefficients as in~\cite{Guevara:2021abz}}
\begin{equation}
    R^{3-2q,a}(z,\zb) = \sum_n\sum_{m=1-q}^{q-1}\frac{s^{q,a}_{m,n}}{(q+m-1)!(q-m-1)!z^{n+2-q}\zb^{m+1-q}}
\end{equation}
the OPE is equivalent to the mode algebra
\begin{equation}
    \left[s^{p,a}_{n,m},s^{q,b}_{n,m}\right] = -if^{ab}_{\ \ c}s^{p+q-1,c}_{n+n',m+m'}.
\end{equation}
A similar formula gives the soft algebra for gravity, the wedge subalgebra of the loop algebra of $w_{1+\infty}$ \cite{Strominger:2021mtt}. Because these soft generators can be obtained by commutators of universal soft theorems, which receive only limited corrections from non-minimal couplings, it is expected that any reasonable celestial dual for Yang-Mills or gravity should contain this algebra in some form.

\subsection{Self-Dual Yang-Mills}
While it would be nice if we could consider the most general theory in the bulk, it is easier to consider some specific cases that are more tractable and serve as sort of toy-models. Self-Dual Yang-Mills (SDYM) theory is a relatively soluble 4D gauge theory where one helicity of the gluon has been projected out, so that the field strength is self-dual. SDYM can be defined by a simple Lagrange multiplier enforcing self-duality of the field strength: 
 \begin{equation}
     S = \int d^4x\Tr B F_-
 \end{equation}
where $F_- = F - *F$ is the anti-self-dual part of the field strength~\cite{Costello:2022wso}. Because the self-duality condition restricts the type of gluons that can propagate in internal lines in Feynman diagrams, SDYM has an extremely simple $\mathcal{S}$-matrix consisting of a tree level $++-$ three-point amplitude and 4-point and higher all $+$ one-loop amplitudes  that are rational functions of the external momenta~\cite{Chalmers:1996rq}. 

SDYM can also be consistently coupled to neutral scalars
\begin{equation}
\label{eq:action}
    S = \int d^4x \left[\Tr BF_- + (-1)^p \phi\square^p\phi - m^2\phi^2 -\frac{1}{2} k\phi \Tr F^2\right].
\end{equation}
Again, the tree-level connected $\mathcal{S}$-matrix is incredibly simple. Because $F = *F$ for the self-dual theory, the Bianchi identity implies that $\nabla_\mu F^{\mu\nu} = 0$ so that the equation of motion for the gauge field does not depend on the scalar coupling. As such, the connected color-ordered amplitudes are the all-gluon amplitudes, which vanish at tree level, and the all-gluon one-scalar amplitude, for which the color ordered amplitude takes the simple form~\cite{Dixon:2004za}
\begin{equation}
    A\left(1^+\cdots n^+(n+1)^0\right) = \frac{2k m^4}{\langle 12 \rangle \langle 23\rangle \cdots \langle n1\rangle}\delta^{(4)}(p_1+p_2+\cdots p_n+p_{n+1}).
\end{equation}
Here, the amplitude does not depend on the power $p$ appearing in \ref{eq:action} because there are no internal scalar exchanges. 

\subsection{Sourced Self-Dual Yang-Mills Amplitudes}
Recent work~\cite{Casali:2022fro,Melton:2022def} has shed light on the importance of thinking about SDYM in the presence of a source. It turns out that SDYM with a scalar source also has a simple $\mathcal{S}$-matrix. If we consider a source for the scalar, with action
\begin{equation}
    S = \int d^4x \left[\Tr BF_- + (-1)^p\phi\square^p\phi - \frac{1}{2}k\phi\Tr F^2 + (-1)^n \eta J(x)\phi(x)\right]
\end{equation}
the tree-level connected all $+$ color-ordered amplitude in momentum space takes the form
\begin{equation}
\label{eq:npts}
    A(1^+\cdots n^+) = \frac{2k\eta (P^2)^2}{\langle 12 \rangle\cdots \langle n1\rangle} \frac{J(P)}{(P^2)^n}
\end{equation}
where $P = p_1 + \cdots p_n$. Because the field strength is restricted to be self-dual, the connected gluon amplitude is exactly proportional to $\eta$, as no connected gluon diagrams exist that can interact with the scalar background more than once and the tree-level all $+$ amplitudes vanish~\cite{Melton:2022def}. Additionally, in the limit where $k \to 0$ with $k\eta = \mu$ fixed, the gluon $\mathcal{S}$-matrix decouples from the dynamical scalar, and we can consider the effective action
\begin{equation}
     S = \int d^4x \left[\Tr BF_-  -\frac{1}{2} \mu\phi(x)\Tr F^2\right]
\end{equation}
where $\phi(x)$ is a fixed function satisfying $\square^n\phi(x) = J(x)$. In this limit, the tree-level $\mathcal{S}$-matrix is derived by simply exponentiating the connected amplitudes given in~\eqref{eq:npts} and subsequently taking $k\eta \to \mu$. This limit projects out amplitudes involving the exchange of the dynamical $\phi$ field, allowing us to replicate pure gluon amplitudes in a scalar background.

\section{Gluon-Gluon OPE Ansatz}
\label{sec:algdual}
The singular gluon-gluon OPE is severely constrained by universal soft and collinear singularities of gauge theory scattering amplitudes. As such, any 2D celestial dual should contain operators that generate the gluon-gluon OPE in gauge theory, possibly with some modifications from the few effective operators that can deform the singular part of splitting function. 

Taking inspiration from~\cite{Stieberger:2022zyk}, we focus on realizations of the gluon-gluon OPEs where a gluon is a product of a level-0 Kac-Moody current $J^a$ and a scalar primary $\Phi_{\Delta-1}^i$, of the form
\begin{equation}\label{eq:opdef}
    \O^{+a}_{\Delta} \equiv \Gamma(\Delta-1)J^a_i\Phi^i_{\Delta-1}.
\end{equation}
where $i$ is some internal index that we have summed over. If the currents form a level-0 Kac-Moody algebra of the form
\begin{equation}
    J^a_iJ^b_j = \frac{-if^{ab}_{\ \ c}\ g_{ij}^{\ \ r}}{z_{12}}J^c_r(z_2) + O(z_{12})
\end{equation}
and the scalars obey the OPE
\begin{equation}
\label{eq:scalope}
    \Phi^i_{\Delta_1}(z_1,\zb_1)\Phi^j_{\Delta_2}(z_2,\zb_2) = h^{ij}_{\ \ r}\Phi^r_{\Delta_1+\Delta_2}(z_2,\zb_2) + O(|z_{12}|)
\end{equation}
where 
\begin{equation}
\label{eq:gheq}
    g_{ij}^{\ \ r}h^{ij}_{\ \ s} = \delta^{r}_{\ \ s}
\end{equation}
these operators will have the OPE.\footnote{While this gives only the $O(\zb_{12}^0)$ term, the subleading terms are fixed by Lorentz invariance.} Note that Equation \ref{eq:gheq} fixes $h$ in terms of $g$. 
\begin{equation}\label{eq:gluonOPE}
\O^{+a}_{\Delta_1}(z_1,\zb_1)\O^{+b}_{\Delta_2}(z_2,\zb_2) = -\frac{\Gamma(\Delta_1-1)\Gamma(\Delta_2-1)}{\Gamma(\Delta_1+\Delta_2-2)}\frac{if^{ab}_{c}}{z_{12}}\O^{+c}_{\Delta_1+\Delta_2-1}(z_2,\zb_2) + \cdots     
\end{equation}
which replicates the gluon-gluon OPE in gauge theory~\cite{Pate:2019lpp}. By taking appropriate residues of this operator product expansion, we can similarly obtain an ansatz for the soft gluon algebra \cite{Guevara:2021abz}:
\begin{equation}
    \begin{split}
        R^{k,a} &= \lim_{\epsilon\to 0}\epsilon \O^{+a}_{k+\epsilon} = \frac{(-1)^{1-k}}{(1-k)!}J^a_i\Phi^i_{k-1} \\
        R^{k,a}(z_1,\zb_1)R^{\ell,b}(z_2,\zb_2) &= -\frac{if^{ab}_{\ \ c}}{z_{12}}\frac{(2-k-\ell)!}{(1-k)!(1-\ell)!}R^{k+\ell-1,c}(z_2,\zb_2) + \cdots.
    \end{split}
\end{equation}

 It should also be noted that such operator constructions are also reminiscent of the ambitwistor string\footnote{We thank Atul Sharma for pointing this out to us.} discussed in~\cite{Adamo:2019ipt}. However, we do not comment on that further here.

\subsection{Extracting Sourced Amplitudes from Realizations of the Gluon Gluon OPE}
While realizing the gluon-gluon OPEs in the boundary theory is suggestive, a true duality will require matching scattering amplitudes to boundary correlators. Since the realization of the gluon-gluon OPE in~\eqref{eq:gluonOPE} contains a centerless Kac-Moody current algebra, correlation functions of these operators will vanish.

To resolve this, we will give the Kac-Moody current algebra a level $k$ so that correlation functions\footnote{This bears resemblance to the way celestial correlators are expected to factorize under soft factorization. In that case, the correlator of currents represents the soft part of the $\mathcal{S}$-matrix while the correlator of, scalars in this case, acts like that of the hard Wilson line dressings. It is also reminiscent of the color-factor term relating color-ordered amplitudes to full amplitudes in gauge theories.}
\begin{eqnarray}
\label{eq:factamp}
    \left\langle \O^{+a_1}_{\Delta_1}(z_1,\zb_1) \cdots \O^{+a_n}_{\Delta_n}(z_n,\zb_n) \right\rangle = \left\langle J^{a_1}_{i_1} \cdots J^{a_n}_{i_n}\right\rangle \left\langle \Phi^{i_1}_{\Delta_1-1}\cdots\Phi^{i_n}_{\Delta_n-1}\right\rangle 
\end{eqnarray}
are nonvanishing. We will describe the physical meaning of this level in Section~\ref{sec:scoup}. For now, we simply use it to extract non-vanishing boundary correlators. We focus on the fully connected contributions from here on. The scalar two point function is constrained by conformal invariance to be 
\begin{equation}
\left\langle \Phi_{\Delta_1}^i(z_1,\bar{z}_1)\Phi_{\Delta_2}^j(z_2,\bar{z}_2)\right\rangle = \frac{\delta_{\Delta_1,\Delta_2}S^{ij}_{\Delta_1,\Delta_2}}{|z_{12}|^{2\Delta_1}}
\end{equation}
where $S^{ij}_{\Delta_1,\Delta_2}$ is some constant in terms of the structure of the operators and their conformal dimensions. Using our ansatz for the gluon operators, 
\begin{eqnarray}
\label{eq:twopointansatz}
    \left\langle \O^{+a}_{\Delta_1}(z_1,\zb_1)\O^{+b}_{\Delta_2}(z_2,\zb_2) \right\rangle &=& \Gamma(\Delta_1-1)\Gamma(\Delta_2-1)\frac{k\delta^{ab}}{z_{12}^2}\left\langle \Phi^i_{\Delta_1-1}(z_1,\zb_1)\Phi^i_{\Delta_2-1}(z_2,\zb_2)\right\rangle \cr
    &=& \frac{k\delta^{ab}\Gamma(\Delta_1-1)\Gamma(\Delta_2-1)\delta_{\Delta_1,\Delta_2}S^{ii}_{\Delta_1-1,\Delta_2-1}}{z_{12}^2|z_{12}|^{2(\Delta_1-1)}}
\end{eqnarray}
where $i$ is summed over. If the space of internal indices is large, we should have $S^{ii}_{\Delta_1,\Delta_2} = N S_{\Delta_1,\Delta_2}$. Comparing with the dependence of the sourced gluon amplitudes implies the strength of the background $\eta$ will equal $N$. Taking $\eta = N \to \infty$ while fixing $\mu = kN$ will decouple the gluons from the dynamical scalar field and lead to finite correlators with vanishing level. 

Taking the soft limit and matching with the two-point amplitude in sourced SDYM \cite{Melton:2022def}
\begin{equation}
    \left\langle R^{j,a}R^{\ell,b}\right\rangle = \frac{\mu\delta^{j\ell}\delta^{ab}a_{-1-j}}{2(-4)^jz_{12}^2|z_{12}|^{2(j-1)}}
\end{equation}
implies that this describes two-point scattering around a scalar background given by
\begin{equation}
\label{eq:backeq}
    \phi(p) =\sum_ja_j(p^2)^j \ \ \mbox{where} \ \ a_j = \frac{2}{(-4)^{1+j}(j+2)!}S_{-2-j,-2-j}.
\end{equation}
Note that we have stripped off the current correlator $\langle J^aJ^b\rangle$. Because $\Phi^i_\Delta$ transforms as a conformal primary of weight $\Delta$, this expression does not depend on $z_{ij},\zb_{ij}$. Thereby, given a realization of the scalar OPE,~\eqref{eq:backeq} gives a background around which gluon two point functions match that computed with the boundary ansatz.

\subsection{General Constraints on Sourced Amplitudes}
While we can always match the two-point function by choosing the background, since we can vary the two-point function for each weight independently by changing the scalar background, consistency with three-point gluon amplitudes provides a constraint on the boundary realization of the scalar operator $\Phi_\Delta$.  
It was shown in~\cite{Melton:2022def} that soft three-point functions of positive helicity gluons scattering around a fixed scalar background\footnote{We assume that the scalar background $\phi$ is damped in the UV and that near $p=0$ it admits a power law expansion in $p^2$.} $\phi(p) = \sum_j a_j(p^2)^j$ take the form\footnote{Here $\binom{n}{n_1,n_2,n_3} = \frac{n!}{n_1!n_2!n_3!}$ is a trinomial coefficient.}
\begin{equation}
\begin{split}
    \A^\phi\left(1^+_{k_1}2^+_{k_2}3^+_{k_3}\right) &= \frac{A_{k_1k_2k_3}}{z_{12}z_{23}z_{31}|z_{12}|^{k_1+k_2-k_3-1}|z_{23}|^{k_2+k_3-k_1-1}|z_{31}|^{k_1+k_3-k_2-1}} \\
   A_{k_1k_2k_3} &= - \frac{a_{-2-\beta/2}}{(-1)^{3\beta/2}2^{\beta+3}}\begin{pmatrix} -\beta/2 \\ \frac{k_1-k_2-k_3+1}{2}, \frac{-k_1+k_2-k_3+1}{2}, \frac{-k_1-k_2+k_3+1}{2}\end{pmatrix}.
\end{split}
\end{equation}
Above, we saw that choosing $\phi(x)$ allowed us to realize different boundary duals for the scalar component of a realization of the gluon-gluon OPE. However, modifying the scalar background does not allow us to set all three-point amplitudes independently. Rather, the ratio of three point amplitudes with different external weights but the same value of $\beta = k_1+k_2+k_3-3$ is fixed. For all soft amplitudes, this implies that
\begin{eqnarray}
    A_{k_1+1,k_2-1,k_3} = \frac{1-k_1+k_2-k_3}{3+k_1-k_2-k_3}A_{k_1k_2k_3}.
\end{eqnarray}
This constrains the three point function of the scalar fields $\Phi_k$, which by conformal invariance has the form
\begin{eqnarray}
    \left\langle \Phi^i_{k_1}\Phi^j_{k_2}\Phi^k_{k_3}\right\rangle = \frac{S^{ijk}_{k_1k_2k_3}}{|z_{12}|^{k_1+k_2-k_3}|z_{23}|^{k_2+k_3-k_1}|z_{31}|^{k_3+k_1-k_2}}.
    \end{eqnarray}
In terms of these three-point functions, the constraint becomes
\begin{eqnarray}
\label{eq:scalarconstraint}
    g_{ijk}S^{ijk}_{k_1+1,k_2-1,k_3} = \frac{(1-k_2)(k_1-k_2+k_3)}{k_1(2+k_1-k_2-k_3)}g_{ijk}S^{ijk}_{k_1,k_2,k_3}.
\end{eqnarray}
where $g_{ijk}$ contains the dependence of the three-point KM correlator on the internal indices. Whether the constraint in~\eqref{eq:scalarconstraint} is ultimately implied by the leading OPE of the scalar field or if it is a novel constraint on the dual scalar theory is left to future work. Additionally, dual theories that fail to satisfy this constraint are not necessarily sick. They merely cannot arise from expanding around a background for a scalar field non-minimally coupled to self-dual Yang-Mills through a $\phi \Tr F^2$ interaction. 

 We now describe two distinct realizations of the gluon-gluon OPE of the form given in~\eqref{eq:opdef} with trivial internal index that can be realized by two different sources for a scalar field.

\section{Gluons from Vertex Operators}
\label{sec:vertex}
We first consider a realization of the scalar OPE~\eqref{eq:scalope} arising from light operators in the semi-classical limit of Liouville theory. While this discussion largely parallels the construction of MHV amplitudes from Liouville theory~\cite{Stieberger:2022zyk}, our scalar source is slightly different since we are looking at all $+$ amplitudes in self-dual Yang-Mills rather than MHV amplitudes. 
\subsection{Vertex Operator Realizations of the Gluon-Gluon OPE}
Realizations of the gluon OPE can also be found using light operators in the large background charge limit of Liouville CFT. The Liouville CFT is defined by the action~\cite{DiFrancesco:639405} 
\begin{equation}
    S = \frac{1}{4\pi}\int d^2x\sqrt{g}\left[\partial^\mu\phi\partial_\mu\phi + Q R \phi + \lambda e^{2b\phi}\right]
\end{equation}
where the background charge is $Q = b + \frac{1}{b}$ and $\lambda$ is the parameter related to the cosmological constant. The spectrum of Liouville theory is continuous, and is given by scalar vertex operators labeled \footnote{Properly, this is a limit of the vertex operator as $\phi \to -\infty$ and the potential vanishes.}
\begin{equation}
    V_\alpha = e^{2\alpha\phi}
\end{equation}
of weight $\Delta_\alpha = 2\alpha(Q-\alpha)$. The semi-classical limit of Liouville theory is the limit $b \to 0$. In this limit, the operators $V_{b\Delta/2}$ have weight $\Delta$ and are the light operators in semi-classical Liouville theory. Operators with momenta that scale as $b^{-1}$ are the hard operators. In the classical limit of Liouville theory, the correlation function of some number of light and heavy operators takes a simple form, as insertions of light operators do not change the saddle of the path integral, while insertions of heavy operators act as a source
\begin{equation}
\label{eq:lightliouville}
    \left\langle V_{b\Delta_1/2}\cdots V_{b\Delta_n/2} V_{\eta_1/b}\cdots V_{\eta_m/b}\right\rangle \approx e^{-S[\phi_c]/b^2}\prod_{k=1}^ne^{\Delta_k\phi_c(z_k,\zb_k)}.
\end{equation}
Here $\phi_c = b\phi(x)$ solves the classical equation of motion with sources created by the hard operators~\cite{BalasubramanianLiouville:2017}
\begin{equation}
    \partial\bar{\partial}\phi_c = 2\lambda e^{\phi_c} - 2\pi\sum_{j=1}^m\eta_j\delta^{(2)}(z-z_i).
\end{equation}
The correlation functions~\eqref{eq:lightliouville} allow us to immediately derive the coincident limit of the light operators. Inserted in a correlation function in the limit $b \to 0$, two light operators simply give the insertion
\begin{equation}
    V_{b\Delta_1/2}(z_1,\zb_1)V_{b\Delta_2/2}(z_2,\zb_2) = e^{\frac{\Delta_1\phi_c(z_1,\zb_1) + \Delta_2\phi_c(z_2,\zb_2)}{2}} = V_{b(\Delta_1+\Delta_2)/2} + O(z_{12}, \zb_{12}).
\end{equation}
As per~\eqref{eq:opdef} we can then define
\begin{equation}\label{eq:liouville}
    \O^{+a}_{\Delta}(z,\zb) = \Gamma(\Delta-1)J^a(z)\lim_{b\to 0}V_{b(\Delta-1)/2}(z,\zb).
\end{equation}
These operators then obey
\begin{equation}
    \O^{+a}_{\Delta_1}(z_1,\zb_1)\O^{+b}_{\Delta_2}(z_2,\zb_2) = -\frac{if^{ab}_{\ \ c}}{z_{12}}B(\Delta_1-1,\Delta_2-1)\O^{+c}_{\Delta_1+\Delta_2-1}
\end{equation}
and realize the gluon-gluon OPE for SDYM which is also the positive helicity gluon OPE in~\cite{Pate:2019lpp}.

\subsection{Celestial Liouville for SDYM}
Taking the gluon operator as defined in~\eqref{eq:liouville}, we obtain the following correlator
\begin{equation}
    \left\langle \O^{+a_1}_{\Delta_1}\cdots\O^{+a_n}_{\Delta_n}\right\rangle = \frac{\mu \Tr[T^{a_1}\cdots T^{a_n}]}{z_{12}z_{23}\cdots z_{n1}} \lim_{b \to 0} \left\langle V_{b(\Delta_1-1)/2}(z_1,\zb_1)\cdots V_{b(\Delta_n-1)/2}(z_n,\bar{z}_n)\right\rangle_b  + \cdots
\end{equation}
where $\cdots$ includes other color orderings and the $\langle \rangle_b$ is a correlator in Liouville theory. Here, we have taken the decoupling limit where the background freezes out. In general the correlation functions have a rather complex structure, however for certain cases they can be written out elegantly. When $\sum_j(\Delta_j-1) = 2$, the Liouville correlator has the form~\cite{Stieberger:2022zyk}
\begin{equation}
\mathcal{S}_n = \lim_{b\to0}\left\langle V_{b(\Delta_1-1)/2}(z_1,\zb_1)\cdots V_{b(\Delta_n-1)/2}(z_n,\bar{z}_n)\right\rangle_b = \int d^2z\prod_{j=1}^n|z_j-z|^{-2(\Delta_j-1)}.
\end{equation}
Using the parametrization of null momenta given above and explicitly imposing the delta function constraint, we can rewrite this integral as a projectivation over the lightcone as\footnote{ Here we have used the integral $\int d\omega\omega^{\beta-1} = 2\pi\delta(\beta)$. This is valid for $\beta \in i\mathbb{R}$ but can be continued off of this region as described in \cite{Donnay:2020guq}.}
\begin{eqnarray}
    2\pi\delta(\beta-2)\mathcal{S}_n &=& 2\pi\delta(\beta-2)\int d^2z\prod_{j=1}^n\left(-\frac{1}{2}\hat{q}(z,\zb) \cdot \hat{q}(z_j,\zb_j)\right)^{1-\Delta_j} \cr
    &=& 4 \int \omega d\omega dzd\zb \prod_{j=1}^n \left(\hat{q}(z_j,\zb_j) \cdot [\omega \hat{q}(z,\zb)]\right)^{1-\Delta_j} \cr
    &=& 2\int d^4X\delta(X^2)\prod_{j=1}^n \left(\hat{q}(z_j,\zb_j) \cdot X\right)^{1-\Delta_j} \cr
    &=& -\frac{2}{\prod_{j=1}^n\Gamma(\Delta_j-1)}\int d^4X\delta(X^2)\prod_{j=1}^n d\omega_j\omega_j^{\Delta_j-2}e^{iX\cdot\sum_{j=1}^n\omega_j\hat{q}(z_j,\zb_j)}.
\end{eqnarray}
Restoring the color part of the gluon operator gives
\begin{eqnarray}
   \left\langle \prod_{j=1}^n\O_{\Delta_j}^{+a_j}(z_j,\zb_j) \right\rangle &=& - \frac{2\mu \Tr[T^{a_1}\cdots T^{a_n}]}{z_{12}\cdots z_{n1}}\int d^4X\delta(X^2)\prod_{j=1}^n d\omega_j\omega_j^{\Delta_j-2}e^{iX\cdot\sum_{j=1}^n\omega_j\hat{q}(z_j,\zb_j)}  + \cdots \cr
    &=& (-2)^{n+1}\mu\int d\omega_j\omega_j^{\Delta_j-1}\frac{\Tr T^{a_1}\cdots T^{a_n}}{\langle 12 \rangle \cdots \langle n1 \rangle}\cr
    & \times & \int d^4X\delta(X^2)e^{iX\cdot\sum_j\omega_j\hat{q}(z_j,\zb_j)} + \cdots
\end{eqnarray}
where $+\cdots$ includes other color orderings. We can recognize this as the connected amplitude describing tree-level gluon scattering around a background for a fourth order scalar with source
\begin{eqnarray}
    J(X) = -\delta(X^2).
\end{eqnarray}
\section{Gluons from Generalized Free Fields}
\label{sec:gff}
Another realization of the soft gluon-gluon OPE can be found by constructing the scalar part of the gluon operator from generalized free fields of weight $-1$. We now show that this realization can be matched, at the level of the two- and three-point functions, by scattering around an exponential background.

\subsection{Generalized Free Field Representations of the soft gluon algebra}
For simplicity we consider the OPE in~\eqref{eq:scalope} for $g = h = 1$. Recall that a generalized free field is one such that the correlation functions are entirely determined from the two point function. This makes it a good simple example so we consider a generalized free-field with OPE 
\begin{equation}
\mu(z_1,\zb_1)\mu(z_2,\zb_2) \sim \gamma|z_{12}|^2.
\end{equation}
Now, consider the scalar operator constructed from these fields as
\begin{equation}
    \Phi_{k} = :\mu^{-k}(z,\zb):
\end{equation}
 where $::$ denotes standard normal ordering and $k$ is negative. Because all terms involving contractions between $\mu$'s are sub-leading in the $z_{12}\zb_{12} \to 0$ limit, they have the OPE 
\begin{eqnarray}
   \Phi_{k}(z_1,\bar{z}_1)\Phi_{j}(z_2,\bar{z}_2) &=& :\mu^{-k}(z_1,\zb_1)::\mu^{-j}(z_2,\zb_2):\cr
   &=& :\mu^{-k-j}(z_2,\zb_2): + O(z_{12},\zb_{12})=\Phi_{k+j}+O(z_{12}).
\end{eqnarray}
Because $\mu$ has negative weight, this matches the ansatz \eqref{eq:scalope} with trivial internal index. As defined in~\eqref{eq:opdef}, we can write a composite operator 
\begin{equation}\label{eq:freefield}
R^{k,a}(z,\zb) := \frac{(-1)^{1-k}}{(1-k)!}J^a(z)\Phi_{k-1}(z,\zb) = \frac{(-1)^{1-k}}{(1-k)!}J^a(z):\mu^{1-k}(z,\zb):    
\end{equation}
that has OPE 
\begin{equation}
    R^{k,a}(z_1,\zb_1)R^{\ell,b}(z_2,\zb_2) = -\frac{if^{ab}_{\ \ c}}{z_{12}}\frac{(2-k-\ell)!}{(1-k)!(1-\ell)!}R^{k+\ell-1,c}(z_2,\zb_2) + \cdots 
\end{equation}
which can be recognized as the soft gluon-gluon OPE.~\cite{Guevara:2021abz}

\subsection{Celestial Generalized Free Fields}
A similar discussion can be had in the case of a generalized free field where the level $k$ is understood to be attached to the Kac-Moody currents. In this case, we can construct non-zero correlation functions. Defining the soft operators as in~\eqref{eq:freefield} we have that, in the decoupling limit,
\begin{equation}
    \left\langle R^{j,a}R^{\ell,b} \right\rangle = \frac{\mu\gamma^{2-2j}\delta^{j\ell}\delta^{ab}}{(1-j)!z_{12}^2|z_{12}|^{2(j-1)}}
\end{equation}
where we have evaluated the two-point function as the maximal contraction of the $\mu$ fields. 

Comparing to the two-point soft-soft amplitude above, the gluon-gluon two point function matches scattering with background 
\begin{equation}
        \phi(p)= \sum_{j=-2}^\infty a_jp^{2j}, \ \ a_j = \frac{2\gamma^{2(2+j)}(-4)^{-1-j}}{(j+2)!}
\end{equation}
Summing this series this gives $\phi(p) = -\frac{8e^{-\gamma^2p^2/4}}{(p^2)^2}$. We can therefore conclude that these amplitudes are generated by coupling self-dual Yang-Mills to a scalar potential with 
\begin{equation}
    \delta S = -\frac{\mu}{2}\int d^4x\phi(x)\Tr F^2
\end{equation}
in the decoupling limit where $\square^2\phi(x) = 4e^{-x^2/\gamma^2}$. 

This generalizes to three-point functions. Computing the three point function of generalized free fields
\begin{equation}
    \left\langle \mu^{n_1}\mu^{n_2}\mu^{n_3}\right\rangle = \frac{C_{n_1n_2n_3}}{|z_{12}|^{n_1+n_2-n_3}|z_{23}|^{n_2+n_3-n_1}|z_{31}|^{n_3+n_1-n_2}}
\end{equation}
as a sum over all possible Wick contractions gives
\begin{equation}
    C_{n_1n_2n_3} = \frac{\gamma^{-\beta/2} n_1!n_2!n_3!}{\left(\frac{n_1+n_2-n_3}{2}\right)!\times\mathrm{cyclic}}
\end{equation}
which satisfies the constraint on the three-point function in~\eqref{eq:scalarconstraint}. This implies that these three-point amplitudes can arise as gluon scattering around a scalar source.

\section{Beyond the scalar decoupling limit}
\label{sec:scoup}
So far we have seen how different choices of scalar field $\Phi$ result in different gluon-gluon OPEs. However, it has been of recent interest~\cite{Ren:2022sws,Melton:2022def} to try to understand deformations to bulk theories that result in specific types of deformations to the boundary OPEs and thereby the corresponding boundary algebras. In particular, one can attempt to understand classes of bulk deformations that result in boundary OPEs that are a consequence of the 2D current $J^a$ satisfying a Kac-Moody algebra with a non-trivial level.

Turning on a level for the WZW current deforms the $J^aJ^b$ OPE to 
\begin{equation}
    J^aJ^b \sim\frac{k\delta^{ab}}{z_{12}^2} - i\frac{f^{ab}_{\ \ c}J^c}{z_{12}}.
\end{equation}
This adds a new term to the OPE of the 2D gluon operators
\begin{eqnarray}
\mathcal{O}_{\Delta_1}(z_1,\bar{z}_1)\mathcal{O}_{\Delta_2}(z_2,\bar{z}_2) & = & \frac{\Gamma(\Delta_1-1)\Gamma(\Delta_2-1)k\delta^{ab}}{z_{12}^2}\Phi_{\Delta_1+\Delta_2-2}\cr
& - & \frac{if^{ab}_c}{z_{12}}\frac{\Gamma(\Delta_1-1)\Gamma(\Delta_2-1)}{\Gamma(\Delta_1+\Delta_2-2)}\mathcal{O}_{\Delta_1+\Delta_2-1}+...
\end{eqnarray}
We see that the OPE of two gluons will now include a scalar term proportional to the level of the current algebra. In the context of a bulk theory, this means we have added an interaction between gluons and a bulk scalar field that now contributes to the Feynman diagram calculations of scattering amplitudes. 

This deforms the soft OPE to 
\begin{equation}
    R^{\ell,a}(z_1,\zb_1)R^{j,b}(z_2,\zb_2) \sim \frac{(-1)^{2-\ell-j}}{(1-\ell)!(1-j)!}\frac{k\delta^{ab}}{z_{12}^2}\Phi_{\ell+j-2}  -\frac{if^{ab}_{\ \ c}}{z_{12}}\frac{(2-\ell-j)!}{(1-\ell)!(1-j)!}R^{\ell+j-1,c}(z_2,\zb_2) + \cdots 
\end{equation}
Defining $\phi_k = \frac{(-1)^{-k}}{(-k)!}\Phi_{k}$, the OPE becomes 
\begin{equation}
    R^{k,a}(z_1,\zb_1)R^{j,b}(z_2,\zb_2) \sim \frac{(2-j-k)!}{(1-k)!(1-j)!}\frac{k\delta^{ab}}{z_{12}^2}\phi_{\ell+j-2}  -\frac{if^{ab}_{\ \ c}}{z_{12}}\frac{(2-\ell-j)!}{(1-\ell)!(1-j)!}R^{\ell+j-1,c}(z_2,\zb_2) + \cdots 
\end{equation}
which is the soft algebra for SDYM coupled to a fourth order scalar with action 
\begin{equation}
    S = \int d^4x \left[\Tr BF_- + \phi\square^2\phi - \frac{1}{2}k \phi \Tr F^2\right] 
\end{equation}
so that the level of the boundary WZW current algebra is dual to a $\phi \Tr F^2$ coupling in the bulk.\footnote{On both sides of this duality, there is an additional $\O(z_{12}^{-1})$ term in the OPE coming from subleading contributions to the gluon-gluon-scalar operator product expansion. These terms do not neatly factorize and may be related to multiparticle operators in celestial CFT. We leave their characterization to future work.}

While we leave determining the precise scalar theory described by the $\phi_k$ operators to future work, it is clear that it must have several interesting and surprising features. The scalar-scalar-scalar operator product expansion, by the usual bulk-boundary dual for OPEs of massless particles~\cite{Himwich:2021dau}, implies that the scalar is interacting, and the theory contains the interaction vertex 
\begin{equation}
S_{int}(\phi) \supseteq \int d^4x \phi \partial^\mu\partial^\nu\phi\partial_\mu\partial_\nu\phi.   
\end{equation}
However, the theory likely contains other interaction terms and must have surprising cancellations in scattering amplitudes: the operator product expansion between two scalars would generically contain a normal ordered term as well as the leading splitting function. For this theory, different diagrams must conspire to cancel the leading multi-particle term in the scalar-scalar OPE. It is likely that the results from recent work on multi-OPEs in celestial CFT~\cite{Ball:2023sdz} would be crucial to understanding this further.

\section{Conclusions and Future Work}
In this work, we have argued that, given a realization of the hard or soft gluon OPE of the form in~\eqref{eq:opdef} obeying the constraint~\eqref{eq:scalarconstraint}, we can find a background for SDYM coupled to a scalar field where gluon scattering amplitudes match those computed by the two-dimensional theory in a decoupling limit. We defined conditions on the three-point function of these scalar operators such that a bulk scalar background exists and described two explicit examples where this is the case. 
 
Additionally, we observed that the level of the boundary Kac-Moody is a coupling constant for a $\phi \Tr F^2$ interaction. When the level is non-zero, bare scalar operators appear in the gluon-gluon operator product expansion as conformally soft modes of a bulk fourth-order axion. This suggests that the scalar part of the gluon operator is the boundary dual to a special bulk scalar field with a particular coupling to the field strength. 

Interestingly, several examples closely related to those studied here have appeared in prior work. The construction of celestial Liouville theory is a close parallel of this story for MHV amplitudes in Yang-Mills theory and obeys the same factorization between a Kac-Moody current and a scalar piece to correct for the weight~\cite{Stieberger:2022zyk}. Additionally, the proposed dual for Burns space holography also has the form of a Kac-Moody current coupled to a field that corrects for the weight, although the Burns space dual is chiral and has a large internal gauge group~\cite{Costello:2023hmi}.

Nevertheless, much work is still to be done. While we have identified the level of the boundary Kac-Moody current algebra as a particular coupling constant in the bulk, the boundary duals we have constructed thus far have no obvious dual to the strength of the background. Interestingly, for SDYM, the strength of the scalar background counts the number of disconnected pieces contributing to a gluon amplitude as each disconnected piece can interact with the background only once. This suggests that the strength of the background should be related to the rank of an internal gauge symmetry and that our decoupling limit is the large $N$ limit of the 2D theory, as it is in Burns space holography~\cite{Costello:2023hmi}. 

Additionally, turning on a level for the boundary Kac-Moody current algebra couples the bulk theory to an interacting fourth-order scalar field, whose characterization we have left to future work. Similar fourth-order scalar theories have appeared in studies of self-dual Yang-Mills on twistor space, where it serves to cancel an anomaly preventing self-dual Yang-Mills from being local on twistor space at the quantum level~\cite{Costello:2022wso}, and in Burns space holography where it appears as Mabuchi gravity~\cite{Costello:2023hmi}. Understanding the scalar theory in this work therefore promises to help construct new complete examples of celestial dualities where the boundary theory has an independent definition. 

Finally, it is important to understand how general the construction of bulk duals to boundary theories is. In this work, we have focused on cases where we tune a source for a scalar field to create non-distributional amplitudes, but there is reason to believe the construction may be more general. In~\cite{Costello:2022wso}, it was shown that conformal blocks of the soft chiral algebra were in a one-to-one correspondence with form factors of the bulk theory, which describe scattering around operators inserted in the bulk. The two-dimensional models we consider replicate this chiral algebra in their singular terms, suggesting that it may be possible to realize any two-dimensional model appropriately realizing the soft algebra of gauge theory by deforming self-dual Yang-Mills with some operator insertion.

It would also be interesting to extend this duality to gravity. While the structures of the $S$-algebra and the $w_{1+\infty}$-algebra are reminiscent of each other, there are several important challenges to overcome. First, the structure constants of the $w_{1+\infty}$-algebra depend less trivially on the weights of the generators, so separating the algebra into a simple `color factor' and scalar part may be more difficult. Additionally, the universal coupling of gravity to matter complicates the construction of the translation-breaking background.

\section*{Acknowledgements}
The authors would like to thank Erin Crawley, Andrea Puhm, Lecheng Ren, Romain Ruzziconi, Andrew Strominger, Tomasz Taylor, and Bin Zhu for interesting discussions. The authors also thank Atul Sharma and Tianli Wang for constructive comments on earlier drafts of this paper. WM is supported by the NSF GRFP grant DGE1745. SN acknowledges support by the Celestial Holography Intitiative at the Perimeter Institute for Theoretical Physics and by the Simons Collaboration on Celestial Holography. SN's research at the Perimeter Institute is supported by the Government of Canada through the Department of Innovation, Science and Industry Canada and by the Province of Ontario through the Ministry of Colleges and Universities.
\bibliographystyle{JHEP}
\bibliography{cpb}
\end{document}